\documentclass{emulateapj}

\newcommand{\tab}[1]{Table~\ref{#1}}
\newcommand{\fig}[1]{Figure~\ref{#1}}

\newcommand{\ngrapes}{1308}    
\newcommand{\nvlt}{81}         
\newcommand{\nall}{114}        
\newcommand{\nfilt}{40}        
\newcommand{\rmsg}{0.046}      
\newcommand{\rmsn}{0.061}      
\newcommand{\zthresh}{1.5}     
\newcommand{\alphaval}{--1.32$\pm$0.07}
\newcommand{\mval}{--22.4$\pm$0.3}

\newcommand{\sex}{\texttt{SExtractor}}
\newcommand{\hyperz}{\texttt{HyperZ}}
\newcommand{\bpz}{\texttt{BPZ}}

\begin{document}
\title{The Galaxy  Luminosity Function at $z\!\simeq\!1$  in the HUDF:
Probing the Dwarf  Population\footnote{Based on observations made with
the NASA/ESA Hubble Space Telescope, obtained from the Data Archive at
the  Space  Telescope Science  Institute,  which  is  operated by  the
Association  of Universities  for Research  in Astronomy,  Inc., under
NASA contract NAS 5-26555.}}

\shorttitle{$B$-band LF at $z\!\simeq\!1$}

\author{R. E. Ryan Jr.\altaffilmark{2}, N. P. Hathi\altaffilmark{2}, S. H. Cohen\altaffilmark{3}, S. Malhotra\altaffilmark{3}, J. Rhoads\altaffilmark{3}, R. A. Windhorst\altaffilmark{3}, T. Budav\'{a}ri\altaffilmark{4}, N. Pirzkal\altaffilmark{5}, C. Xu\altaffilmark{5}, N. Panagia\altaffilmark{6}, L. Moustakas\altaffilmark{7}, S. di Serego Alighieri\altaffilmark{8}, \& H. Yan\altaffilmark{9}}

\altaffiltext{2}{Department of Physics, Arizona State University, Tempe, AZ 85287}
\altaffiltext{3}{School of Earth and Space Exploration, Arizona State University, Tempe, AZ 85287}
\altaffiltext{4}{Department of Physics and Astronomy, The Johns Hopkins University, Baltimore, MD 21218}
\altaffiltext{5}{Shanghai Institute of Technical Physics, 500 Yuitan Rd., Shanghai, P.R. China 200083}
\altaffiltext{6}{Space Telescope Science Institute, Baltimore, MD 21218}
\altaffiltext{7}{JPL/Caltech, 4800 Oak Grove Dr., MS 169-327, Pasadena, CA 91109}
\altaffiltext{8}{INAF-Osservatorio Astrofisico di Arcetri, Largo Enrico Fermi 5, I-50125 Firenze, Italy}
\altaffiltext{9}{California Institute of Technology, MS 100-22, Pasadena, CA 91125}

\email{russell.ryanjr@asu.edu}

\shortauthors{Ryan Jr. et al}
\begin{abstract}

We   present   a   catalog   of  spectro-photometric   redshifts   for
\ngrapes~galaxies from the GRism ACS Program for Extragalactic Science
(GRAPES) observations  with the  {\it Hubble Space  Telescope}.  These
low-resolution   spectra   between   6000~\AA\   and   9500~\AA\   are
supplemented with  $U$, $J$, $H$,  and $K_s$ from  various facilities,
resulting  in  redshifts computed  with  $\sim\!40$~spectral bins  per
galaxy.   For \nvlt\  galaxies  between $0.5\!<\!z\!<\!\zthresh$  with
spectroscopic  redshifts,  the standard  deviation  in the  fractional
error in $(1+z)$ is \rmsg.  With this catalog, we compute the $B$-band
luminosity function in this redshift range from 72 galaxies.  Owing to
the depth  of the GRAPES survey,  we are able  to accurately constrain
the   faint-end   slope   by   going  to   $M_B\!\simeq\!-18$~mag   at
$z\!=\!1.0\!\pm\!0.2$,  nearly two  magnitudes  fainter than  previous
studies.  The  faint-end slope is  $\alpha\!=$\alphaval. When compared
to  numerous published  values at  various redshifts,  we  find strong
evidence for a  steepening of the faint-end slope  with redshift which
is expected in the hierarchical formation scenario of galaxies.

\end{abstract}


\section{Introduction} \label{introduction}

Measuring a large sample of accurate redshifts for distant galaxies is
one  of  the  most  daunting  tasks  in  observational  cosmology  and
extragalactic science.  Given  the typical brightness (AB$\sim$26~mag)
of distant  ($z\!\gtrsim\!1$) galaxies, optical  spectroscopy requires
extensive  observations on the  largest telescopes.   While multi-slit
spectrographs  and   grating  prism  (grism)  modes   allow  for  many
simultaneous spectroscopic observations,  they can be severely limited
in wavelength coverage and spectral resolution.  Therefore, the use of
photometric redshifts  estimated from  observed fluxes is  a necessity
for   the   statistical   study   of   distant   galaxies   \citep[for
example][]{music,coe,mob06}.

Since   the    release   of    the   original   Hubble    Deep   Field
\citep[HDF;][]{hdf},  photometric  redshifts have  been  the focus  of
numerous studies and are the  cornerstone of many others.  At present,
there  are  primarily  two   different,  yet  similar  techniques  for
computing   these  redshifts:   $\chi^2$  minimization   and  Bayesian
statistics.  The  minimization scheme compares  a set of  model fluxes
measured  from empirical  or synthetic  spectral  energy distributions
(SEDs)  to the  observed fluxes.   The earliest  mature  studies using
$\chi^2$ minimization were met with some skepticism regarding possible
degeneracies between redshift and  internal reddening in galaxy colors
\citep{lyf}.  With  the addition  of near-infrared ($JHK$)  imaging in
the  HDF, \citet{fly}  showed that  many  of the  degeneracies can  be
broken to yield fairly accurate redshifts ($\sigma[\Delta z/(1+z_{{\rm
spec}})]\!\approx\!0.1$).   In contrast, the  Bayesian marginalization
uses prior redshift probabilities, obtained by other means, to compute
extremely  accurate   redshifts  \citep{bayes}.   While   the  current
implementation  of this technique,  \bpz, produces  reliable redshifts
from limited observational constraints, it does not compute many useful
quantities  \citep[eg.    {\it  k}-corrected  magnitudes,  probability
densities, $V$-band extinction, and age;][]{caputi04}.

The luminosity  function (LF) of galaxies  represents the distribution
of galaxy  luminosities in a redshift  interval.  Since the  LF can be
used to constrain galaxy formation models, it has been well studied at
both  high and  low redshift.   While the  LF can  be computed  at any
wavelength, it  is often studied  in the rest-frame  $B$-bandpass.  At
$z\!\simeq\!1$  accurate determinations  of the  $B$-band  LF requires
large, deep surveys  \citep[][]{chen,gdds,cross,zucca}.  In this work,
we complement these studies  by going nearly two~magnitudes fainter in
$M_B$ and  accurately determining the  shape of the LF  in particular,
the faint-end slope.

Where  necessary, we  assume the  {\it Wilkinson  Microwave Anisotropy
Probe}  cosmology  \citep[WMAP;][]{wmap},  where  $\Omega_0\!=\!0.26$,
$\Omega_{\Lambda}\!=\!0.74$,  and $H_0\!=\!73$~km~s$^{-1}$~Mpc$^{-1}$.
All magnitudes quoted herein are in the AB system \citep{abmag}.  This
paper   is  organized   as  follows:   the  data   are   described  in
\S~\ref{observation},  the details  of the  redshift  measurements are
given in  \S~\ref{spz}, the  redshift quality and  luminosity function
are  in  \S~\ref{results}, and  a  discussion  of  the results  is  in
\S~\ref{discuss}.

\section{Observations} \label{observation}
The     GRism     ACS     Program    for     Extragalactic     Science
 \citep[GRAPES;][]{grapes} data  consists of  40 orbits with  the {\it
 Hubble  Space  Telescope}  (HST)   of  the  Hubble  Ultra-Deep  Field
 \citep[HUDF;][]{beck}   taken   during   Cycle~12.   These   slitless
 spectroscopic  data were taken  with the  ACS in  the G800L  mode and
 range    from   $\sim$5500--10500~\AA\    with   a    resolution   of
 $R\!\simeq\!100$  and   cover  $\simeq\!11$~arcmin$^2$.   The  GRAPES
 observations  were taken  over  four epochs,  each  with a  different
 position  angle in order  to minimize  the contamination  from nearby
 objects.   These  data  were  supplemented with  the  existing  grism
 observations  of  \citet{riess04}, giving  a  total  dataset of  five
 position  angles and  an  integration time  of $1.1\times10^5$~s.   A
 thorough discussion  of the GRAPES observations,  data reduction, and
 spectral extraction and calibration can be found in \citet{grapes}.

\subsection{Additional Bandpasses}

Given the  limited spectral  range of the  GRAPES observations,  it is
necessary to  include the existing  broad-band data available  in this
highly   observed  field   to  increase   accuracy  of   the  redshift
measurements.  To  extend our spectra  to $\sim\!3000$~\AA--2.3$\mu$m,
we include  the CTIO-MOSAIC II, $U$-band  observations, the HST-NICMOS
$J$-  and  $H$-band  \citep{thomp},  and  VLT-ISAAC  $K_s$-band  data.
Therefore, the final  dataset has $\sim$40~independent spectral points
from the combination of broad-band and grism observations.

Since the $i'$-band image is  the deepest optical exposure ever taken,
it is used to define  the apertures for the near-ultraviolet (NUV) and
near-infrared (NIR)  data.  To eliminate  aperture corrections between
the broad-band data, we convolve  all images to the same full-width at
half-maximum as  the worst observation  (the $U$-band data),  which is
$1\farcs3$.  While  convolving the  HST images is  highly undesirable,
the NUV constraint is  essential to accurately determine the redshifts
at $1\!\lesssim\!z\!\lesssim\!2.3$  \citep{ferg}. \citet{stan} suggest
that  surveys  with $U$-  and  $B$-band  data  can expect  drastically
reduced catastrophic failures for galaxies, while quasar colors remain
degenerate  and the  redshifts may  be unreliable.   Since the  PSF is
$\simeq\!20$  times  larger for  the  ground-based observations,  many
objects  in the  HST  images become  confused  after the  convolution.
Objects were deemed confused if after the convolution $>$1 object from 
the original GRAPES catalog \citep{grapes} was within the new aperture.
Therefore,  $U$-band flux  was only  measured for  galaxies  that have
unconfused detections, the remaining  galaxies were only measured from
the unconvolved images and do not have a $U$-band measurement.

The final  photometric dataset includes \ngrapes\  galaxies with grism
spectroscopy.  Since the HST-NICMOS data covers $\sim$50\% of the HUDF
observed  with  ACS, 687/\ngrapes\  galaxies  have  $J$- and  $H$-band
imaging  and only  543/\ngrapes\ galaxies  have {\it  viable} $U$-band
data.   In total,  273/\ngrapes\  galaxies have  the  entire suite  of
$UB$+grism+$JHK_s$  data.  All broad-band  magnitudes are  measured as
\texttt{MAG\_AUTO} by \sex\ in dual-image mode with the default values
for   \texttt{PHOT\_AUTOPARAMS}  \citep{sexref}.   To   determine  the
apertures,       we        use       \texttt{DETECT\_THRESH}       and
\texttt{ANALYSIS\_THRESH} of 1.5.

\subsection{Source Catalogs}

For  the matched  aperture photometry,  we use  the  convolved HST-ACS
$i'$-band  as the  {\it detection  image} in  \sex.  Since  the GRAPES
spectra were extracted in a  trace of fixed width \citep{grapes}, they
must be scaled to reproduce the  fluxes measured by \sex\ in the HUDF.
Therefore,   we  define  a   multiplicative  aperture   correction  as
$\beta\!=\!10^{-0.4\left(i'_{{\rm  HUDF}}-i'_{{\rm  GRAPES}}\right)}$,
where  $i'_{{\rm HUDF}}$  and  $i'_{{\rm GRAPES}}$  are the  $i'$-band
magnitudes  measured from  the  ACS imaging  and  the GRAPES  spectra,
respectively.   Since  the $i'$-band  image  was  used  to define  the
apertures used  in the other images,  there is no  need for additional
corrections between the broad-band images.

\begin{figure}
\epsscale{1.1}
\plotone{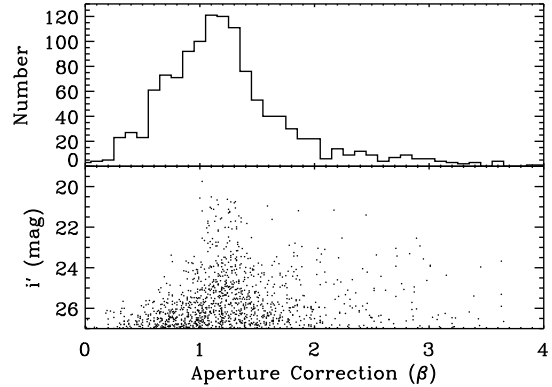}
\caption{Aperture corrections for the $\ngrapes$~GRAPES galaxies.  The
aperture correction is defined  such that the GRAPES spectra reproduce
the $i'$-band  fluxes measured from the  HUDF.  Since the  NUV and NIR
fluxes  are measured  with apertures  defined by  the  $i'$-band data,
there are no additional corrections between the broad-band data.  {\it
Upper panel}--- The distribution  of aperture corrections.  The GRAPES
spectra are only scaled to reproduce the $i'$-band fluxes, since their
usable portion  does not  completely cover the  neighboring bandpasses
($V$  and/or  $z'$).   {\it  Lower  panel}--- The  dependence  of  the
$i'$-band magnitude  on the aperture correction.  The  majority of the
objects  with  $\beta\!\leq\!1$  are   near  the  detection  limit  of
$i'\!\simeq\!27$~mag.   This   suggests  that  uncertainties   in  the
background subtraction may be  the cause of these aperture corrections
\citep{grapes}.  }\label{aperplot}
\end{figure}

The   distribution  of   these  aperture   corrections  is   shown  in
\fig{aperplot}.   The width  of this  distribution is  related  to the
properties  of  the   spectral  extraction,  contamination  of  nearby
objects, and  the broad-band apertures.  Since the  GRAPES spectra are
most  reliable from $\sim$6000-9500\AA,  they do  not fully  cover the
$V$- or  $z'$-bands, therefore  scaling to multiple  bands or  using a
wavelength-dependent  aperture correction is  not possible  with these
data.  Lastly,  the scaled GRAPES  spectra are rebinned onto  a common
wavelength  grid  and are  assigned  a  100~\AA\  wide top-hat  filter
transmission curve.  This procedure  typically decreases the number of
GRAPES  spectral   points  by  a  factor  $\sim$2   and  boosts  their
signal-to-noise,  which  markedly   helps  for  the  faintest  sources
($i'\!\simeq\!27$~mag).

\section{Spectro-Photometric Redshifts}\label{spz}

Measuring  redshifts  from  a  calibrated  spectrum  requires  readily
identifiable  absorption or emission  lines or  characteristic breaks.
Since  the GRAPES  data  have a  resolution  of $R\!\simeq\!100$,  the
narrow    spectral   features    are    typically   not    detectable.
\citet{egrapes} measured emission line redshifts from $\sim$100~GRAPES
galaxies;   therefore,  we   complement  that   study   by  estimating
spectro-photometric  redshifts  for  the  remaining  dataset  from  the
low-resolution GRAPES spectra.

\begin{figure}
\epsscale{1.1}
\plotone{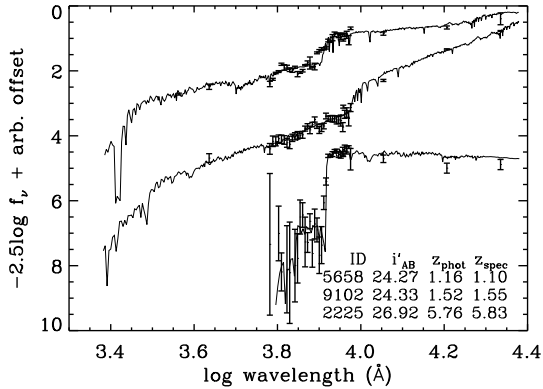}
\caption{Representative SEDs.  These examples highlight two advantages
of  the  GRAPES  data:  high  signal-to-noise  at  faint  flux  levels
($i'\!\sim\!25$~mag) and excellence in determining spectral breaks.  In
the  lower-right, we tabulate  the HUDF  ID, $i'$-band  magnitude, the
spectro-photometric redshift  derived in this work,  and the published
spectroscopic  redshift.   The  spectra  are  arranged  in  descending
redshift so that from top to  bottom the IDs are 5658, 9102, and 2225.
With  the addition  of the  Wide-Field Camera~3  (WFC3)on HST,  we can
expect   UV  prism   and  IR   grism  data   at   1900--4500~\AA\  and
1.1--1.7~$\mu$m,  respectively.   The  combination  of  ACS  and  WFC3
spectra  will   provide  an  extremely  broad   spectral  coverage  at
low-medium resolution.  }\label{repseds}
\end{figure}

From   the  scaled   spectra   and  broad-band   fluxes,  we   compute
spectro-photometric     redshifts    with    the     code,    \hyperz\
\citep{hyperzref}.  The alternative code, \bpz\ \citep{bayes}, was not
used for  three reasons:  It currently does  not constrain  the galaxy
ages to be less than the age of the Universe, it does not compute many
useful, secondary  quantities, and  with our spectral  constraints and
coverage, \bpz\  is not expected to  substantially outperform \hyperz.
\citet{mob04}  compares the  quality  of redshifts  from \hyperz\  and
\bpz\  for  a  sample  of  $\simeq$400~galaxies in  the  Chandra  Deep
Field-South using  broad-band fluxes.  They show that  the rms scatter
in  the photometric  redshifts is  0.140  and 0.072  for $\chi^2$  and
Bayesian  techniques, respectively.  In  \fig{repseds}, we  show three
observed  SEDs  with  best-fit  templates where  there  is  reasonable
agreement  between  our  redshift  and  the  published  value.   These
examples  illustrate the  power of  the GRAPES  spectra to  locate the
4000~\AA\  break (the  top two)  and the  Ly$\alpha$-break  (the lower
one).

\begin{figure}
\epsscale{1.1}
\plotone{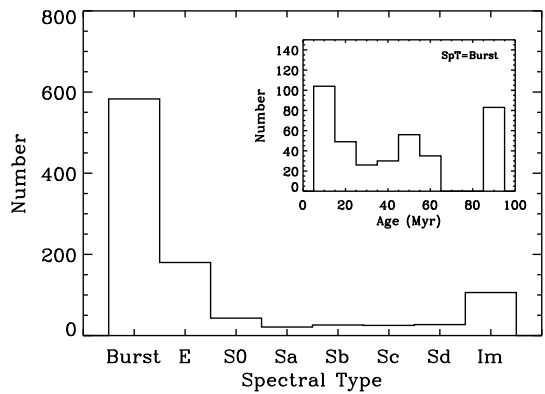}
\caption{The distribution  of BC03  spectral types.  Clearly  the {\it
Burst} template is the dominant SED and from the inset, these galaxies
are  typically  very  young.   This  is somewhat  expected  since  the
majority of  galaxies at $z\!\sim\!1$  are young and  actively forming
stars.   Since the  BC03  templates  are tabulated  as  a function  of
stellar population age, we require that  this age be less than the age
of the Universe at the given redshift.}\label{sptplot}
\end{figure}

Using  the  set  of  SED  templates  summarized  in  \tab{hyperztemp},
\hyperz\  minimizes  the reduced  $\chi^2$  between  the observed  and
modeled fluxes  as a  function of redshift,  age, and  extinction.  By
neglecting  the  $Vi'z'$-bands  in  the  spectro-photometric  redshift
calculation, the  combination of  scaled GRAPES spectra  and remaining
broad-band fluxes  typically yields $\simeq\!\nfilt$~fully independent
spectral bins per galaxy.   In \fig{sptplot}, we show the distribution
of  best-fit spectral  types for  the \ngrapes\  galaxies.   Since the
\citet[BC03;][]{bc03} templates  are generated  at a series  of finite
time   steps  of  $\Delta\log{(t/{\rm   1~Gyr)}}\!\lesssim\!0.05$,  we
require the galaxy to be less than  the age of the Universe at a given
redshift. The  {\it Burst} template  is a single,  instantaneous burst
followed  by passive stellar  evolution.  Since  this template  is the
most  common, we  show the  the distribution  of stellar  ages  in the
inset.  We see that majority of the {\it Burst} models are very young,
which is  not surprising, since galaxies  are expected to  be blue and
actively forming stars at $z\!\sim\!1$ \citep[eg.][]{egrapes}.

\begin{table}
\caption{\hyperz\ Templates}
\label{hyperztemp}
\begin{tabular*}{0.48\textwidth}
  {@{\extracolsep{\fill}}lr}
\hline
\hline
\multicolumn{1}{c}{SpT\tablenotemark{$\dagger$}} & \multicolumn{1}{c}{$\tau$\tablenotemark{$\ddagger$}}\\
$ $&$ $\\
\hline
1&0\\
2&1\\
3&2\\
4&3\\
5&5\\
6&15\\
7&30\\
8&$\infty$\\
\hline
\tablenotetext{$\dagger$}{All of the templates came from the BC03 stellar population synthesis models assuming an exponential star formation history.}
\tablenotetext{$\ddagger$}{$\tau$ is the $e$-folding time in the exponential function.}
\end{tabular*}
\end{table}

\section{Results} \label{results}

\subsection{Redshift Quality} \label{zqual}
\begin{figure}
\epsscale{1.1}
\plotone{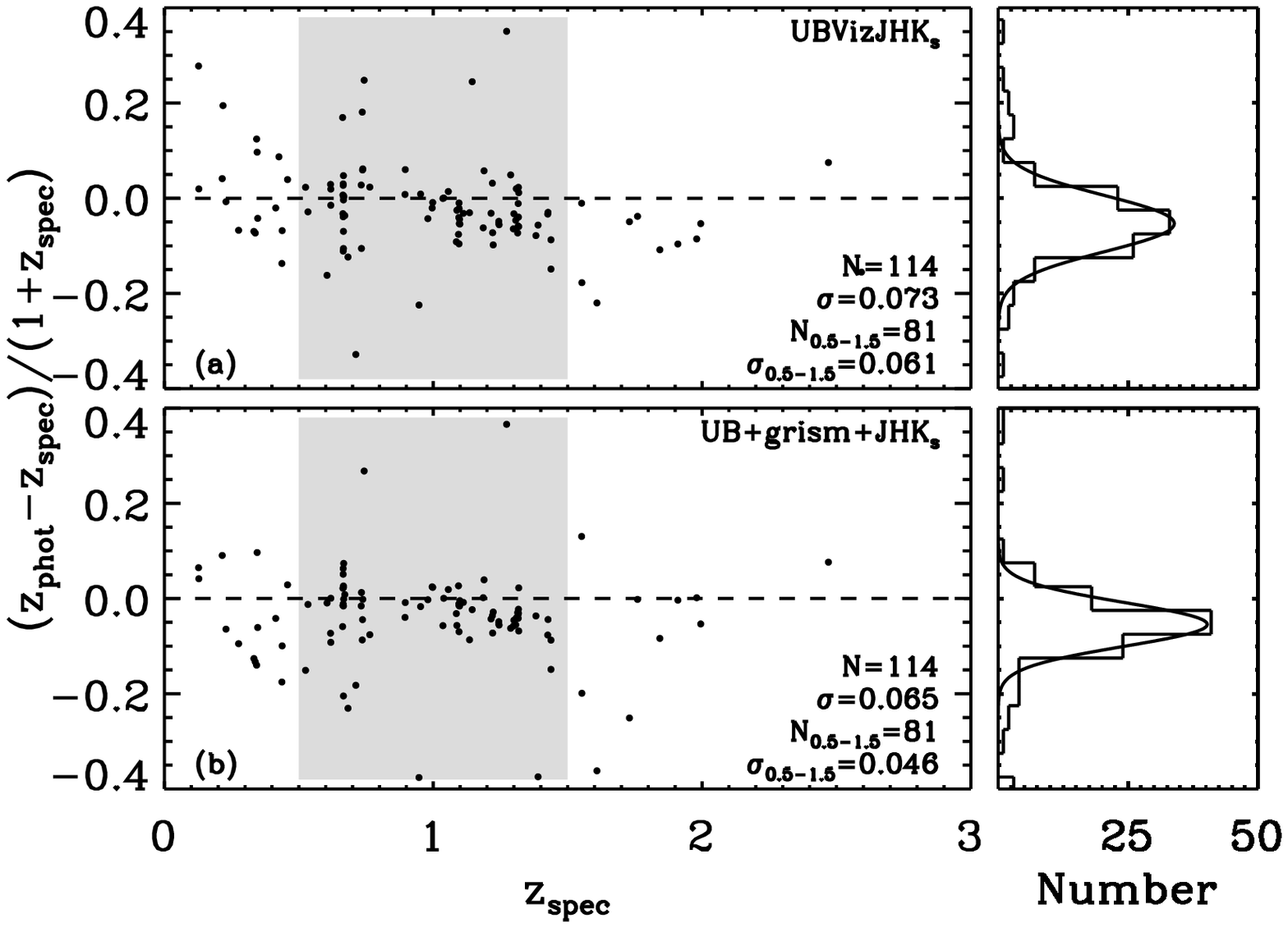}
\caption{Comparison    between    GRAPES    spectro-photometric    and
spectroscopic  redshifts  from  the VLT  \citep{music,ravikumar,vltz}.
The  schemes   for  computing  the  redshifts  is   indicated  in  the
upper-right.   We tabulate  the standard  deviation of  the fractional
error     in    $(1+z)$    for     the    shaded     redshift    range
$0.5\!\leq\!z\!\leq\!1.5$,  when  the 4000~\AA\  break  occurs in  the
grism spectra,  and the  number of galaxies  in the  lower-right.  The
distribution of  the residuals is  shown to the  right of each  of the
primary  panels.  Seven  catastrophic outliers  have  been eliminated,
since they  contained strong emission lines  \citep{egrapes} which are
not accounted  for in the stellar population  synthesis models.  While
the broad-band data alone can  yield good redshift estimates, with the
low-resolution  grism  spectra  we  further refine  the  redshifts  to
$\sigma_{0.5-1.5}\!=\!\rmsg$.  Since these data constitute the extreme
case of  the best possible  data (ultra-deep ACS and  NICMOS imaging),
the redshift refinement  is only marginal.  However, we  expect a much
more  substantial improvement in  the more  general case  of shallower
imaging without a broad spectral coverage.  }\label{compzplot}
\end{figure}

We  compare   our  spectro-photometric  redshifts   to  the  published
spectroscopic   measurements  \citep{music,ravikumar,vltz}.    Of  the
\ngrapes\  GRAPES galaxies,  only \nall\  have  measured spectroscopic
redshifts.   For  these galaxies,  we  show  the  fractional error  in
$(1+z)$   as   a   function   of   the   spectroscopic   redshift   in
\fig{compzplot}.   The  upper and  lower  panels  show  the error  for
different sets of fluxes used in the redshift computation as indicated
in  the  upper right  corners.   Photometric  redshifts  are the  most
reliable when a readily identifiable spectral break occurs between two
adjacent bandpasses.   The 4000~\AA\ and  Ly$\alpha$-breaks will occur
in    the   GRAPES    spectra   for    $0.5\!\leq\!z\!\leq\!1.5$   and
$4\!\leq\!z\!\leq\!8$, respectively.  Therefore,  when either of these
breaks occur in the GRAPES spectra, we expect accurate redshifts.  The
standard  deviation of the  fractional error  is \rmsn\  for redshifts
computed using only the broad-band observations.  However, when we use
the  low resolution  grism  in  place of  the  $Vi'z'$ bandpasses  the
overall redshift uncertainty reduced to \rmsg\ for the \nvlt\ galaxies
with  $0.5\!\leq\!z\!\leq\!1.5$.  Since  these low  resolution spectra
eliminate  the need  for  deep  $V$ and  $z'$  imaging, this  approach
requires $\sim$10\%  the observing time while  producing more accurate
redshifts and  providing more spectral  information.  In \tab{repcat},
we show  a representative  portion of the  final catalog.   The entire
ascii-readable version is available online.

\subsection{Luminosity Functions} \label{lfs}

We   calculate    the   LF   using   the    $V_{{\rm   max}}$   method
\citep{schm68,will97}  in 1.0~mag wide  bins from  absolute magnitudes
measured from  \hyperz\ by convolving  the best-fit spectrum  with the
rest-frame  $B$-bandpass.   We  construct  three  redshift  intervals:
$z\!=\!1.0\pm0.2$  which has  the best  statistics, $z\!=\!0.9\pm0.1$,
and $z\!=\!1.1\pm0.1$ for direct  comparison to previous studies.  The
differential  LF  is  then  computed  by  $\Phi(M)dM\!=\!N(M)dM/\Delta
V(z_{{\rm  max}})$, where  $N(M)dM$  is the  number  of galaxies  with
absolute magnitudes between $M$ and $M+dM$, and
\begin{equation}
\Delta V(z_{{\rm max}})=\frac{\Omega}{4\pi}\int_{z_{{\rm min}}}^{z_{{\rm max}}}C(z,M)\frac{dV}{dz}dz,
\end{equation}
where  $\Omega$ is  the solid  angle  of the  GRAPES survey,  $z_{{\rm
max}}$  is the maximum  redshift at  which a  galaxy with  an absolute
magnitude  of  $M_B$ would  have  been  detected  in the  survey,  and
$C(z,M_B)$  is  the   completeness  function.   The  uncertainties  on
$\Phi(M)dM$ are computed assuming Poisson noise from the galaxy counts
\citep[eg.][]{wolf03}.

To estimate the  completeness in each absolute magnitude  bin we use a
Monte  Carlo  simulation  \citep{wolf03,tamas}.   First,  we  generate
random  redshifts from a  Gaussian distribution  of width  $\rmsg$, as
expected from \fig{compzplot}.  Next,  we count the number of deviates
with apparent magnitudes brighter than  the survey limit, and take the
ratio   of  number   recovered  to   the  number   simulated   as  the
completeness. This  correction is typically  less than a factor  of 10
for the absolute magnitudes presented.

\begin{figure}
\epsscale{1.1}
\plotone{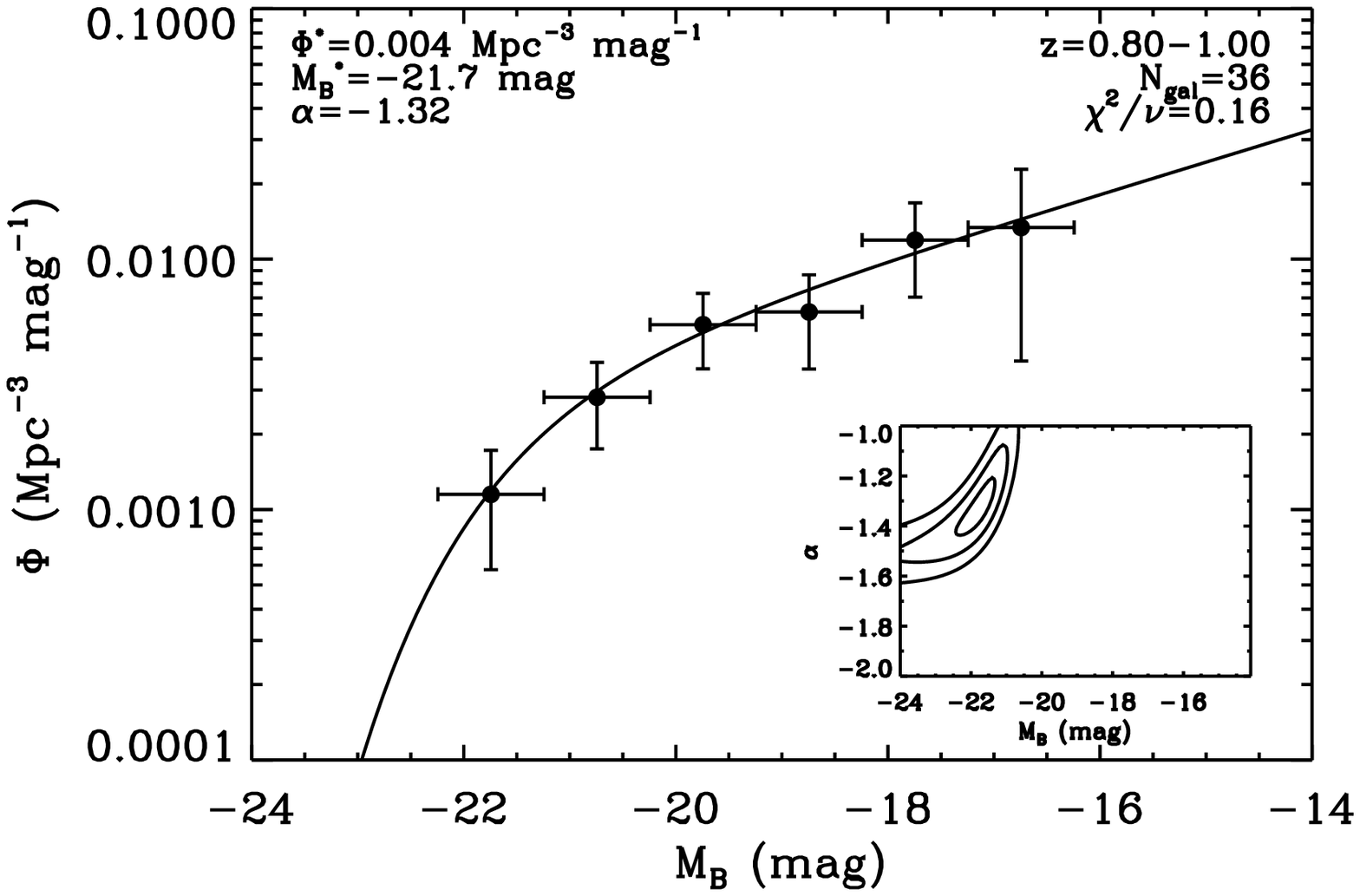}
\plotone{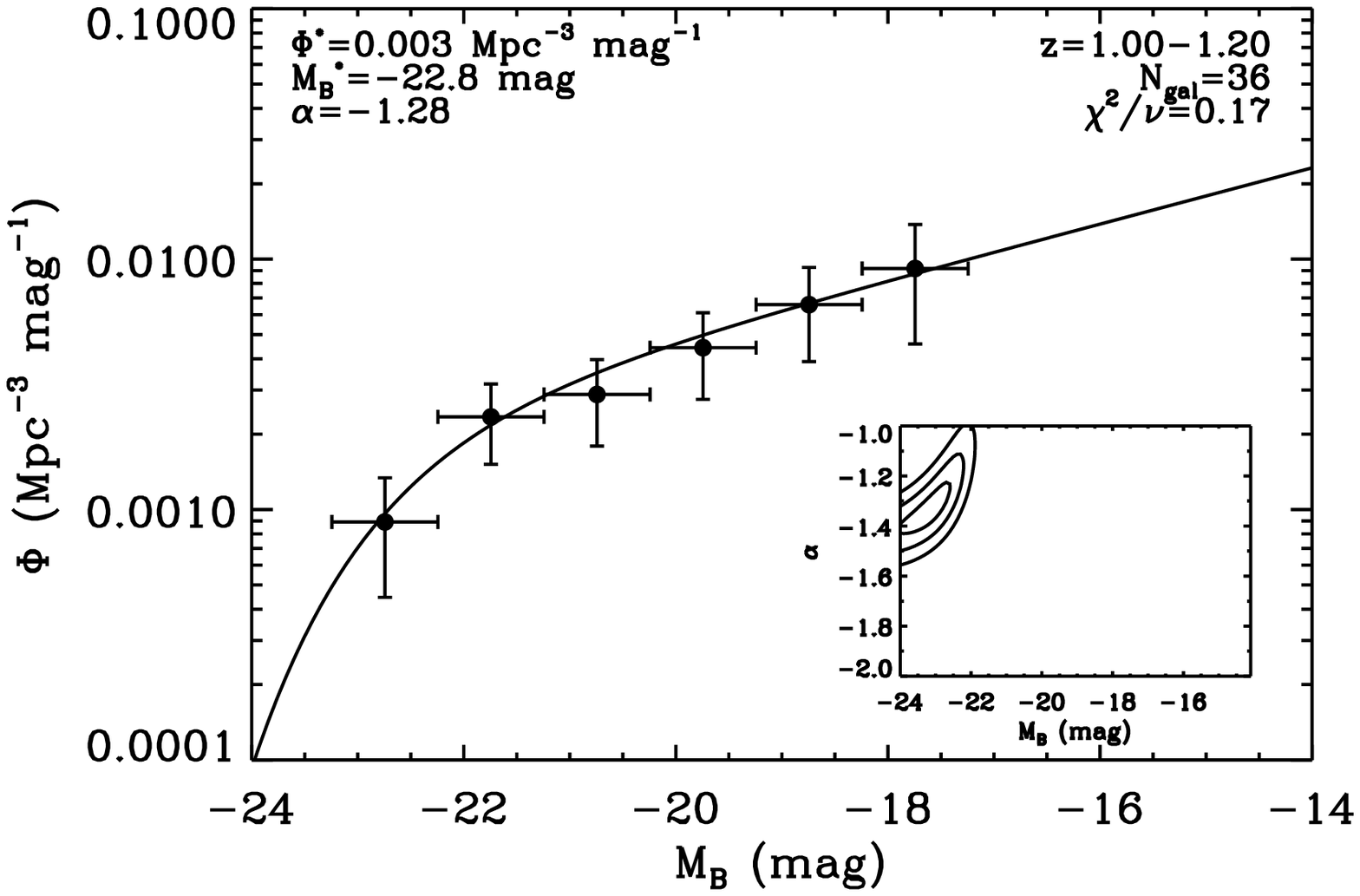}
\caption{The two $B$-band luminosity functions for $z\!=\!0.8\!-\!1.0$
(top)   and  $z\!=\!1.0\!-\!1.2$   (bottom)  with   the  corresponding
$\Delta\chi^2_{\nu}$  contours in  the ($\alpha\!-\!M^*$)  plane.  The
offset  and scatter in  \fig{compzplot} will  only alter  the absolute
magnitudes by $\Delta M\!\simeq\!0.1$~mag at these redshifts. Owing to
the faint flux limit of  the GRAPES survey ($i'\!=\!27.2$~mag), we are
able to  directly determine the Schechter parameters  in this critical
redshift  interval  for  galaxy   evolution.   While  these  data  can
accurately  determine  $M^*\!=$\mval\  and $\alpha\!=$\alphaval\  (for
$z\!=\!1.0\!\pm\!0.2$), the normalization  ($\Phi^*$) is less certain,
since  the object  selection and  contamination in  grism spectroscopy
poses  significant  challenges.   Therefore,  the union  of  complete,
ground-based surveys  with these deep grism  observations provides the
most thorough results.}\label{lumfuncs}
\end{figure}

In \fig{lumfuncs}, we show  the LFs for $z\!=\!0.90\!\pm\!0.10$ (left)
and  $z\!=\!1.10\!\pm\!0.10$  (right).   We  model these  LFs  with  a
standard Schechter function  \citep{sche76}, which is parameterized by
the   normalization  ($\Phi^*$),  characteristic   absolute  magnitude
($M^*$),  and   the  faint-end  slope   ($\alpha$).   Additionally  in
\fig{lumfuncs},  we show the  contours for  $\Delta\chi_{\nu}^2$=1, 4,
and  9 in  the ($\alpha\!-\!M^*$)  plane as  insets.  While  the total
number   of   galaxies   in   these   redshift  bands   may   be   low
($\sim\!10$~galaxies per  absolute magnitude bin),  the GRAPES dataset
provides  excellent  constraints  of  the  Schechter  parameters  (see
\tab{schpar}).   Previous   studies  at  these   redshifts  often  use
ground-based  spectroscopic surveys  which are  inherently  limited to
$M_B\!\sim\!-19$~mag    \citep[such   as   ][]{chen,gdds,cross,zucca}.
Consequently, it has been customary  to {\it assume} a faint-end slope
of $\alpha\!\simeq\!-1.3$  \citep[eg.][]{will06}.  However, the GRAPES
observations provide a  means to {\it measure} the  faint-end slope of
$\alpha\!=$\alphaval\ at $z\!=\!1.0\!\pm\!0.2$.  It is reassuring that
the  assumption of  $\alpha\!\simeq\!-1.3$ at  these redshifts  is not
wholly incorrect.

\begin{table}
\caption{Best Fit Schechter Parameters}
\label{schpar}
\begin{tabular*}{0.48\textwidth}
   {@{\extracolsep{\fill}}lrccc}
\hline
\hline
\multicolumn{1}{c}{$z$} & \multicolumn{1}{c}{$N_{{\rm gal}}$} & \multicolumn{1}{c}{$\Phi^*$} & \multicolumn{1}{c}{$M_B^*$} & \multicolumn{1}{c}{$\alpha$}\\
\multicolumn{1}{c}{$ $} & \multicolumn{1}{c}{$ $} & \multicolumn{1}{c}{(10$^{-4}$ Mpc$^{-3}$ mag$^{-1}$)} & \multicolumn{1}{c}{(mag)} & \multicolumn{1}{c}{$ $}\\
$ $&$ $&$ $&$ $&$ $\\
\hline
0.90$\pm$0.10&36&35.9$\pm$0.2&--21.7$\pm$0.9&--1.32$\pm$0.19\\
1.10$\pm$0.10&36&25.8$\pm$0.1&--22.8$\pm$0.5&--1.28$\pm$0.10\\
1.00$\pm$0.20&72&26.1$\pm$0.1&--22.4$\pm$0.3&--1.32$\pm$0.07\\
$ $&$ $&$ $&$ $&$ $\\
\hline
\end{tabular*}
\end{table}

\subsection{Redshift Evolution of the Faint-end Slope}\label{aofz}

\begin{figure}
\epsscale{1.1}
\plotone{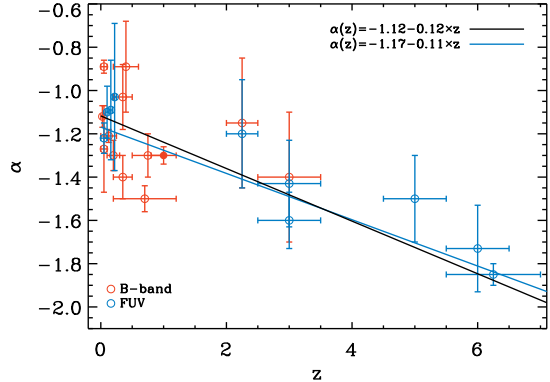}
\caption{A compendium of the faint-end slope for the galaxy luminosity
function.  We show 23~published slopes from 18~sources, including this
work.  While  there are  many more values,  we require each  survey to
have  measured the  LF $\sim$2~mag  fainter than  $M^*$  and optimally
selected   their   galaxies  to   minimize   the  {\it   k}-correction
\citep{ilbert}  to  ensure  uniform  and  reliable  estimates  of  the
faint-end slope.  The red points indicate the measurements made in the
rest-frame                                                     $B$-band
\citep{lin97,saw97,mar98,fri01,nor02,bla03,dri03,wolf03,mar07},     the
blue       points       represent       the       rest-frame       FUV
\citep{ste99,iwa03,yan,tamas,wyd05,bou06,saw06},  and  the filled  red
point  at $z\!=\!1.0\!\pm\!0.2$  is  the result  of  this work.   This
suggests that  dwarf galaxies were  more numerous at high  redshift as
predicted in the hierarchical formation paradigm.}\label{alphaevol}
\end{figure}

The hierarchical formation scenario states that many dwarf galaxies at
high  redshift  will  merge  over  cosmic time,  which  results  in  a
increased  number of  dwarf galaxies  at high  redshift.   This effect
could  be  observed  as  a  steepening of  the  faint-end  slope  with
redshift.  In \fig{alphaevol}, we  show the redshift dependence of the
faint-end  slope  compiled   from  numerous  studies.   These  studies
include,  but are not  limited to,  the largest  ground-based redshift
surveys     \citep[eg.][]{nor02,bla03},     deepest    HST     surveys
\citep[eg.][]{beck},       and       nearby,       Galex       surveys
\citep[eg.][]{tamas,wyd05}.   While each  survey has  unique selection
effects and  observational biases, many  problems can be  mitigated by
requiring two  criteria of  the dataset: a  sufficient amount  of data
{\it   fainter}  than   $M^*$   and  a   minimal   reliance  on   {\it
k}-corrections.  \citet{ilbert} emphasize  that only surveys for which
$(1+z_{{\rm    obs}})\!\sim\!\lambda^S/\lambda^{{\rm   REF}}$   (where
$z_{{\rm  obs}}$  is the  redshift  of  the  LF, and  $\lambda^S$  and
$\lambda^{{\rm REF}}$  are the wavelengths  at which the  galaxies are
selected and the LF is  computed, respectively) can reliably infer the
faint-end slope.   When the \citet{ilbert} condition is  met, the {\it
k}-correction  is minimized. For  uniformity, we  only show  values of
$\alpha$ which  meet these requirements.  In  \fig{alphaevol}, the red
and blue filled circles indicate  the faint-end slopes measured in the
rest-frame  $B$-band and  far UV  (FUV$\simeq$1700~\AA), respectively.
Our  GRAPES  observation  is  indicated  as an  open,  red  square  at
$z\!=\!1.0\!\pm\!0.2$.

From \fig{alphaevol}, the faint-end  slope clearly depends on redshift
in the  manner suggested in  the hierarchical scenario:  many low-mass
galaxies at  high redshift which,  throughout cosmic time,  merge into
the massive galaxies of today.  We parametrize the redshift dependence
of the faint-end slope as  $\alpha(z)\!=\!a+bz$, and give the best fit
models for  the entire dataset  and the FUV  as black and  blue lines,
respectively in  \fig{alphaevol}.  While  the two fits  are consistent
with each other and the observations, the significant scatter warrants
further study.  There are many  effects which could contribute to this
scatter:  differences  in  the  rest-frame  bandpass,  type  dependent
evolution,   galaxy   clustering   and  large-scale   structure,   and
non-uniform sample selections.

\section{Discussion} \label{discuss}

We present  a catalog  of \ngrapes\ spectro-photometric  redshifts for
the objects observed  in the GRAPES project.  When  the GRAPES spectra
are  supplemented  with $UJHK_s$  fluxes  from  other facilities,  the
standard  deviation  of  the  fractional  error  in  $(1+z)$  for  the
\nvlt~GRAPES     galaxies     in     the     redshift     range     is
$0.5\!\leq\!z\!\leq\!\zthresh$  is  \rmsg.    While  this  is  only  a
marginal  improvement  over   traditional  photometric  redshifts,  it
requires  $\sim$10\% exposure  time.  Since  the  photometric redshift
technique is  essentially a ``break-finding''  algorithm, the redshift
accuracy  of  the  GRAPES  observations  is limited  by  the  spectral
resolution    of   the    grism.    Therefore,    we    can   estimate
spectro-photometric redshifts at a  comparable depth and accuracy with
only $\simeq$50 orbits with HST  (40 orbits for the grism observations
and 10 orbits  for an $i'$-band exposure for  object selection).  This
is  a  critical  improvement  for  wide angle  surveys  which  may  be
completely  reliant on  traditional photometric  redshifts.  Moreover,
the  grism observations provide  a necessary  complement to  the deep,
ground-based spectroscopic surveys.  Owing to the line-spread function
and  resolution  of  the  grism spectra,  determining  redshifts  from
emission lines, in a  fashion similar to ground-based observations, is
typically  not  possible.   However,  from  the  high  signal-to-noise
continua, the  spectral breaks  at flux levels  not possible  from the
ground can provide excellent redshift measurements.

Since the  HST-ACS grism  observations allow for  significantly deeper
spectral observations,  we are  able to measure  the faint-end  of the
$B$-band luminosity function at  redshifts not currently possible from
the  ground.  In  the  hierarchical formation  scenario, galaxies  are
expected to evolve by successive merging over cosmic time.  Therefore,
we expect to find fewer dwarf galaxies at lower redshifts which can be
measured in terms of the evolution of the Schechter parameters.  These
data are able to measure  the faint-end slope in the critical redshift
range of $z\!=\!0.5\!-\!1.5$, where  the cosmic star formation rate is
substantially  changing.   When our  faint-end  slope  is compared  to
numerous  studies, we  find strong  evidence for  a redshift-dependent
$\alpha$.   While  previous authors  have  suggested  a similar  trend
\citep[eg.][]{arn05,zucca},  the   compilation  of  published  results
provides both increased statistics and redshift range.

The redshift dependence of the other two Schechter parameters has been
discussed  by   other  authors.   \citet{lin99}   study  the  redshift
evolution of  the $B$-band  galaxy LF, and  in particular  propose the
parameterizations         of         $M^*(z)\!=\!M^*(0)-Qz$        and
$\rho(z)\!=\!\rho(0)10^{0.4Pz}$,  where  $\rho\!=\!\int\Phi(M)dM$.   If
$\alpha$  is constant  with  redshift, then  $\rho$  and $\Phi^*$  are
essentially equivalent.  These parameters ($P,Q$) provide a simple way
of quantifying galaxy evolution and can be determined as a function of
galaxy     type.      For     $q_0\!=\!0.1$,    \citet{lin99}     find
$(P,Q)\!=\!(-1.00\pm0.40,1.72\pm0.41)$  for the  combination  of early
and  intermediate type  galaxies.  \citet{fri01}  parametrize $\Phi^*$
linearly  on redshift:  $\Phi^*(z)=a+b(1+z)$,  where there  is only  a
minimal dependence  of galaxy type on the  coefficients ($a,b$).  From
the  observed  redshift-evolution  of  the  Schechter  parameters,  an
obvious  trend is  emerging:  the galaxy  LF  is shallower  and has  a
brighter characteristic absolute magnitude at low redshift.

The   improvement  with  the   grism  spectroscopy   over  traditional
photometric redshifts on a  {\it per observation basis} bears strongly
on the future  NASA missions, such as the  Wide-Field Camera~3 upgrade
for the {\it  Hubble Space Telescope} and the  planned {\it James Webb
Space Telescope}.

\acknowledgments  We   thank  the  anonymous  Referee   for  the  many
suggestions which improved the  paper. We acknowledge the support from
the Arizona State University NASA  Space Grant (to RER). This work was
supported  by grants GO  9793 and  GO 10530  from the  Space Telescope
Science Institute,  which is operated  by AURA under NASA  contract NAS
5-26555.

\begin{table}
\caption{Representative\tablenotemark{$\dagger$} Catalog}
\label{repcat}
\begin{tabular*}{0.99\textwidth}
   {@{\extracolsep{\fill}}lrrcccccccc}
\hline
\hline
\multicolumn{1}{c}{HUDF} & \multicolumn{1}{c}{$x$} &\multicolumn{1}{c}{$y$} & \multicolumn{1}{c}{RA} & \multicolumn{1}{c}{Dec} & \multicolumn{1}{c}{$i'$} & \multicolumn{1}{c}{$\beta$} & \multicolumn{1}{c}{$z_{\rm phot}$} & \multicolumn{1}{c}{SpT} & \multicolumn{1}{c}{Age} & \multicolumn{1}{c}{$M_{B}$}\\
\multicolumn{1}{c}{ID\tablenotemark{*}} & \multicolumn{1}{c}{(pix)} & \multicolumn{1}{c}{(pix)} & \multicolumn{1}{c}{(h:m:s)} & \multicolumn{1}{c}{(\degr\,\arcmin\,\arcsec)} & \multicolumn{1}{c}{(mag)} & \multicolumn{1}{c}{$ $} & \multicolumn{1}{c}{$ $} & \multicolumn{1}{c}{$ $} & \multicolumn{1}{c}{(Gyr)} & \multicolumn{1}{c}{(mag)}\\
$ $&$ $&$ $&$ $&$ $&$ $&$ $&$ $&$ $&$ $&$ $\\
\hline
    1& 4932.88&  802.89&3:32:39.7&--27:49:42.5&22.931$\pm$0.012& 1.14&0.38&5& 7.500&--17.75\\
    5& 5089.39&  753.56&3:32:39.4&--27:49:44.0&26.983$\pm$0.110& 0.60&0.69&1& 0.090&--14.36\\
    7& 5052.73&  788.94&3:32:39.5&--27:49:42.9&25.367$\pm$0.030& 1.21&0.61&1& 0.004&--16.96\\
    8& 5013.85& 1274.58&3:32:39.5&--27:49:28.4&21.464$\pm$0.003& 1.14&0.64&1& 1.434&--20.18\\
   13& 5108.60&  918.70&3:32:39.3&--27:49:39.1&24.985$\pm$0.042& 1.76&0.52&1& 0.360&--16.62\\
$ $&$ $&$ $&$ $&$ $&$ $&$ $&$ $&$ $&$ $&$ $\\
\hline
\tablenotetext{*}{Identification from \citet{beck}}
\tablenotetext{$\dagger$}{The full ascii version is available online at http://wwwgrapes.dyndns.org/}
\end{tabular*}
\end{table}

\end{document}